\documentclass[%
 reprint,
 amsmath,amssymb,
 aps,
prd,
floatfix,
]{revtex4-2}

\usepackage{placeins}
\usepackage{graphicx}% Include figure files
\usepackage{dcolumn}% Align table columns on decimal point
\usepackage{bm}% bold math
\usepackage{hyperref}% add hypertext capabilities

\usepackage{amsmath}
\usepackage{amssymb}
\usepackage{amsfonts}
\usepackage{accents}
\usepackage{algorithm}
\usepackage{algpseudocode}
\usepackage{float}



\newboolean{usePNG}
\setboolean{usePNG}{true}

\def\g{\gamma}

\def\p{\partial}
\def\ptlb{\left(\p_t- \mathcal{L}_\beta\right)}
\def\lm{\mathcal{L}_m}

\def\n{\nabla}
\def\e4c{e^{-4\chi}}
\def\ep4c{e^{4\chi}}

\def\gh{\hat{\gamma}}
\def\Gh{\hat{\Gamma}}
\def\Lh{\hat{\Lambda}}
\def\Rh{\hat{R}}
\def\Ah{\hat{A}}
\def\Dh{\hat{\Delta}}

\def\dh{\hat{D}}
\def\jh{\hat{j}}

\def\no{\accentset{\circ}{\nabla}}
\def\Do{\accentset{\circ}{D}}
\def\go{\accentset{\hphantom{i}\circ}{\g}}

\def\Go{\accentset{\circ}{\Gamma}}

\def\eps{\accentset{\sim}\epsilon}

\def\lp{\left}
\def\rp{\right}

\def\vp{\vphantom{}}

\def\sF{\vp^\star F}

\begin{document}

\preprint{APS/123-QED}

\title{Universality in the Critical Collapse of the Einstein-Maxwell System}

\author{Gray D. Reid}
\affiliation{Department of Physics and Astronomy,
     University of British Columbia,
     Vancouver BC, V6T 1Z1 Canada}

\author{Matthew W. Choptuik}
\affiliation{
     Department of Physics and Astronomy,
     University of British Columbia,
     Vancouver BC, V6T 1Z1 Canada}

\date{\today}% It is always \today, today,

\begin{abstract}
We report on critical phenomena in the gravitational collapse of the 
electromagnetic field in axisymmetry using cylindrical coordinates. We 
perform detailed numerical simulations of four families of dipole and 
quadrupole initial data fine-tuned to the onset of black hole formation. It 
has been previously observed that families which bifurcate into two on-axis 
critical solutions exhibit distinct growth characteristics from those which 
collapse at the center of symmetry. In contrast, our results indicate 
similar growth characteristics and periodicity across all families of initial 
data, including those examined in earlier works. More precisely, for all 
families investigated, we observe power-law scaling for the maximum of the 
electromagnetic field invariant ($\mathrm{max}|F_{\mu\nu}F^{\mu\nu}| \sim 
|p-p^{\star}|^{-2\g}$) with $\g \approx 0.149(9)$. We  find evidence of 
approximate discrete self-similarity in near-critical time evolutions with a 
log-scale echoing period of $\Delta \approx 0.62(8)$ across all families of 
initial data. Our methodology, while reproducing the results of prior 
studies up to a point, provides new insights into the later stages of 
critical searches and we propose a mechanism to explain the 
observed differences between our work and the previous calculations.

\end{abstract}


\maketitle

\section{\label{em_sec:introduction}Introduction}

In this paper we report results from an investigation of critical 
collapse in the Einstein-Maxwell (EM) system, a model where the 
electromagnetic field is coupled to the general relativistic 
gravitational field. We start with a brief review of black hole 
critical phenomena in gravitational collapse, and direct 
those unfamiliar with the subject to the comprehensive review articles 
\cite{carsten1999critical} and \cite{gundlach_critical_phenomena_2007}.

When studying the critical collapse of a gravitational system, we consider 
the evolution of a single parameter family of initial data with the parameter 
$p$ chosen such that when $p$ is sufficiently small, the gravitational 
interaction is weak. As the magnitude of $p$ is increased, the gravitational
interaction becomes strong and for sufficiently large $p$, the time evolution 
of the  system eventually results in a spacetime containing a black hole. By 
carefully tuning $p$, we find a critical parameter $p^{\star}$ representing 
the threshold of black hole formation for that particular family of initial 
data. The behaviour of solutions arising in the near critical regime, 
$p \rightarrow p^{\star}$, is complex and varied; its study comprises the core 
of what is referred to as  critical phenomena in gravitational collapse.  
 
Depending on the particulars of the model, we may find behaviour such as 
the existence of universality in the critical solutions, the scaling of 
physical quantities as functions of $| p - p^{\star}|$, or symmetries of 
the critical solution beyond those imposed by the initial data or model. 
Here, we are exclusively interested in type II critical phenomena which 
was first studied in the context of the collapse of a massless scalar field in 
spherical symmetry~\cite{choptuik_universality_and_scaling_1993}. 

Type II critical phenomena are typically seen in systems with massless or 
highly relativistic matter fields. For these systems, the critical point 
$p^{\star}$ partitions the phase space of solutions in two such that 
for $p < p^\star$ we have complete dispersal while for  $p > p^\star$ the 
final state of the system contains a black hole: the critical solution, 
which is transient and represents neither dispersal nor black hole formation, 
sits at the interface of these two regions. Most studies of type II critical 
phenomena have been performed in the context of spherical symmetry, and until 
stated otherwise we will restrict attention to the spherically symmetric case.

A fundamental property of all type II critical solutions which have been 
determined to date is that they are self similar. Depending on the specific 
matter content of the system under consideration, the critical solution may 
be either continuously self similar (CSS) or discretely self similar (DSS). 
For a CSS spacetime in coordinates adapted to the symmetry, the metric 
coefficients take the form
\cite{gundlach_critical_phenomena_2007}:
\begin{align}
   g_{ab}\lp(\tau, x^i\rp) = e^{-2\tau} \accentset{\sim}{g}_{ab}\lp(x^i\rp),
\end{align}
where $\tau$ is the negative logarithm of a spacetime scale and $x^i$ are 
generalized dimensionless angles about the critical point. For DSS spacetimes 
in adapted coordinates we have instead \cite{gundlach_critical_phenomena_2007}:
\begin{align}
   g_{ab}\lp(\tau, x^i\rp) &= e^{-2\tau} \accentset{\sim}{g}_{ab}
   \lp(\tau,x^i\rp),
\\
\label{em_dss_time_scaling}
   \accentset{\sim}{g}_{ab}\lp(\tau,x^i\rp) &= \accentset{\sim}{g}_{ab}
   \lp(\tau + \Delta,x^i\rp),
\end{align}
where $\accentset{\sim}{g}_{ab}$ is function of $\tau$ and $x^i$ which is 
periodic in $\tau$ with period $\Delta$. Therefore, in the vicinity of 
$p^\star$, a DSS critical solution exhibits periodic scale invariance in 
length and time. In almost all cases which have been studied in spherical 
symmetry the critical solutions which have been found (for both types of 
self-similarity) are universal, by which we mean that they do not depend on 
the specifics of the initial data families that are used to generate 
them~\cite{gundlach_critical_phenomena_2007, matt_global_structure_choptuik, 
koike1995_radiation_fluid, gundlach1997understanding}. The echoing period, 
$\Delta$, when it exists, is similarly universal.

For systems with a CSS critical solution, invariant dimensionful quantities, 
such as the mass of the resulting black hole in the supercritical regime, 
scale according to
\begin{align}
   \label{em_css_mass_scaling}
   \ln{\lp(M\rp)} &= \gamma\ln\lp|p-p^\star\rp|  + c_{\mathrm{M}},
\end{align}
where $\gamma$ is a universal exponent and $c_M$ is some family-dependent 
constant. When the critical solution is DSS, a universal periodic function, 
$f_M$, with  period $\Delta$ is superimposed on this basic power law 
\cite{gundlach_critical_phenomena_2007}:
\begin{align}
   \label{em_dss_mass_scaling}
   \ln{\lp(M\rp)} &= \gamma\ln\lp|p-p^\star\rp| 
   + f_{\mathrm{M}}\lp(\gamma\ln\lp|p-p^\star\rp|\rp) + c_{\mathrm{M}} \, .
\end{align}
Other dimensionful quantities scale in a corresponding manner. For example,
if we were to look at the maximum energy density, $\rho_{\rm max}$,
encountered during a given subcritical simulation (performed in coordinates 
adapted to the self similarity) we would have
\begin{align}
   \label{em_dss_rho_scaling}
   \ln{\lp(\rho_{\rm max}\rp)} &= 
   -2\gamma\ln\lp|p-p^\star\rp| + f_{\rho}\lp(\gamma\ln\lp|p-
   p^\star\rp|\rp) + c_{\rho},
\end{align}
where $f_\rho$ is another universal periodic function and $c_\rho$ is 
another family-dependent constant. Although type II critical solutions are 
generically unstable, they tend to be minimally so: they typically have a 
single unstable mode in perturbation theory and, in the above scaling 
laws, $\gamma$ turns out to be the inverse of the Lyapunov exponent of 
this unstable mode.

Since the original spherically-symmetric scalar field work, many other 
models have been thoroughly investigated.
Going beyond spherical symmetry, among the most important studies are those of 
the critical collapse of axisymmetric vacuum gravitational waves, originally 
examined by Abrahams and Evans~\cite{abrahams_critical_behaviour_1993,
abrahams_universality_1994}. The study of vacuum critical collapse provides 
a means of achieving arbitrarily large space-time curvatures outside of a 
black hole through purely gravitational processes. In the critical limit this 
culminates in the formation of a naked singularity, which continues to be an 
object of great theoretical interest. 
Fundamentally, although the critical features are not unique to the vacuum 
case, the vacuum provides the most natural gravitational context and is 
therefore most likely to provide information relevant to the studies of 
quantum gravity and cosmic censorship.

Simulations of vacuum critical collapse have proven to be difficult 
and replication (or otherwise) of early results has been challenging. 
It has only been in the past few years that work in this context has seen 
significant progress~\cite{ledvinka_slicing_conditions_2018, 
ledvinka_universality_of_curvature_invariants_2021, hilditch2013collapse, 
hilditch2017evolutions, fernandez2022evolution, baumgarte2023critical, 
baumgarte2023criticalhexa}. In particular, advances in formalisms and in the 
choices of gauge has enabled groups to expand upon the original work of 
Abrahams and Evans. 
In general, investigations into the collapse of non-spherically 
symmetric systems have yielded far more complicated pictures than their 
spherically symmetric counterparts, with family-dependent scaling and 
splitting of the critical solution into distinct loci of collapse appearing 
in a number of models~\cite{gundlach_critical_phenomena_2007, 
choptuik2003_msf_axi, ledvinka_universality_of_curvature_invariants_2021, 
abrahams_critical_behaviour_1993, abrahams_universality_1994, 
mendoza_critical_phenomena_electromagnetic_2021, 
baumgarte_critical_phenomena_electromagnetic_2019, 
baumgarte2023criticalhexa, baumgarte2023critical}.

Turning now to the EM system, we note that, as in the case of the pure 
Einstein vacuum, the model has no dynamical freedom in spherical symmetry 
and must therefore exhibit non-spherical critical behaviour. Recently, 
Baumgarte et 
al.~\cite{baumgarte_critical_phenomena_electromagnetic_2019}, and  Mendoza and
Baumgarte~\cite{mendoza_critical_phenomena_electromagnetic_2021} 
investigated the critical collapse of the EM model in axisymmetry. Using a 
covariant version of the formalism of Baumgarte, Shapiro, Shibata and 
Nakamura (BSSN) in spherical polar coordinates, they found evidence for 
family-dependent critical solutions for dipole and quadrupole initial data. 
Specifically, for each type of initial data, distinct values of $\gamma$ and 
$\Delta$ were found.

In this paper, we present the results of our own investigation into the 
critical collapse of the EM system, also in axisymmetry, but using cylindrical 
coordinates. We incorporate an investigation of the critical behaviour in the 
well-studied massless scalar field model to test and calibrate our code, as 
well as to verify the validity of our analysis procedures which are then 
applied to the more complicated EM system.

We investigate a total of five families of initial data for the EM model, 
three of which are new, and the other two which are chosen in an attempt 
to replicate the experiments of Mendoza and
Baumgarte~\cite{baumgarte_critical_phenomena_electromagnetic_2019, 
mendoza_critical_phenomena_electromagnetic_2021}. In contrast to the prior 
work, which yielded distinct scaling exponents for the quadrupolar 
computations relative to the dipolar ones, we find evidence of universality 
in $\gamma$ and $\Delta$  across all families. We do not, however, observe 
evidence for universality in the periodic functions $f_{i}$ as defined in 
(\ref{em_dss_mass_scaling})--(\ref{em_dss_rho_scaling}).

For dipolar-type initial data we find that the collapse occurs at the center 
of symmetry (in this case the coordinate origin) and that the EM fields 
maintain a roughly dipolar character throughout the collapse process. 
Conversely, for quadrupolar initial data, we observe that the system 
eventually splits into two well separated, on-axis, centers of collapse. That 
is, after the initial data is evolved for some period of time, the matter 
splits into two distributions of equal magnitude, each centered on the 
$\rho=0$ axis, with one distribution centered at positive $z$ and the other 
at a corresponding location below the $z=0$ plane. After this bifurcation 
occurs, the matter continues the process of collapse. In the limit that 
$p \rightarrow p^{\star}$, the matter collapses at the points 
$(z,\rho) = (z_c, 0)$ and $(z,\rho) = (-z_c, 0)$; these are the points at 
which a naked singularity would form in the critical limit and we refer to 
them as accumulation points.  Although the evolution of quadrupole initial 
data prior to the bifurcation is initially consistent 
with~\cite{mendoza_critical_phenomena_electromagnetic_2021}, subsequent 
collapse at the mirrored centers appears to become dominated by a critical 
solution which exhibits similar properties to the dipole cases.

\section{\label{em_sec_background}Background}

Our investigation is restricted to the case of axial symmetry. In terms of 
Cartesian coordinates $(x,y,z)$ we adopt the usual cylindrical coordinates, 
$(z,\rho,\phi)$:
\begin{align}
        z    &= z,
        \\
        \rho &= \sqrt{x^2 + y^2},
        \\
        \phi &= \arctan{\left(\frac{y}{x}\right)}.
\end{align}

For both the generation of initial data and its eventual evolution, we limit 
our investigation to the case of zero angular momentum and adopt the line 
element,
\begin{align}
   ds^2 &= \lp(-\alpha^2 + \rho^2\beta_\rho\beta^\rho + \beta_z\beta^z \rp)dt^2
   \\ \nonumber
   &\hphantom{=} 
   +\lp( 2\beta_z dz + 2\rho \beta_\rho d\rho \rp) dt
   \\ \nonumber 
   &\hphantom{=} 
   +\lp( G_a dz^2 + G_b d\rho^2 + \rho^2G_cd\phi^2 
   + 2 \rho G_d dzd\rho  \rp),
\end{align}
with corresponding metric,
\begin{align}
   \label{em_metric}
   g_{\mu\nu} &= 
   \begin{pmatrix}
   -\alpha^2 + \beta_l\beta^l & \beta_j
   \\
   \beta_i                    & \gamma_{ij}
   \end{pmatrix},
   \\ \nonumber
   &=
   \begin{pmatrix}
   -\alpha^2 + \rho^2\beta_\rho\beta^p + \beta_z\beta^z & \beta_z & 
   \rho \beta_\rho & 0
   \\
   \beta_z          & G_a       & \rho G_d  & 0 
   \\
   \rho\beta_\rho   & \rho G_d  & G_b       & 0
   \\
   0                & 0         & 0         & \rho^2 G_c 
   \end{pmatrix}.
\end{align}
Here and below, all spacetime functions, $g$, have coordinate dependence
$g(t,z,\rho)$. For convenience in our numerical calculations and derivations, 
we have chosen the form of the metric components in~({\ref{em_metric}) so that 
all of the basic dynamic variables satisfy
\begin{align}
   \lim_{\rho \rightarrow 0} g(t,z,\rho) = g_0(t,z) + \rho^2 g_2(t,z)
    + \dots
\end{align}
Thus, all of the dynamical variables have even character about 
$\rho=0$. Using standard definitions of the spatial stress tensor $S_{ij}$ 
(with spatial trace $S)$, momentum, $j^i$, and energy density $\rho_E$,
we have
\begin{align}
   S_{ij} &= \g^\alpha \vp_i \g^\beta \vp_j T_{\alpha\beta},
   \\
   S &= \g^{ij}S_{ij},
   \\
   j^{i} &= -\g{^{ij}}\g^\mu \vp_j n^{\nu} T_{\mu \nu},
   \\
   \rho_E &= n^\mu n^\nu T_{\mu \nu} \, .
\end{align}

We adopt the generalized BSSN (GBSSN) decomposition of 
Brown~\cite{brown2009covariant, alcubierre2011formulations, 
sanchis2014fully, daverio2018apples, baumgarte2013numerical} 
and take the so-called Lagrangian choice for the evolution of 
the determinant of the conformal metric, 
\begin{align}
   \p_t \gh &= 0,
\end{align}
such that the equations of motion are given by
\begin{align}
      \lm &= \ptlb,
\\
\label{em_maxwell_lm_chi_bssn}
      \lm \chi &= -\frac{1}{6} \alpha K + \frac{1}{6}  \dh_m\beta^m,
\\
\label{em_maxwell_lm_K_bssn}
      \lm K &= -D^2\alpha + \alpha\lp( \Ah_{ij} \Ah^{ij} 
      + \frac{1}{3} K^2 \rp) 
      \\ \nonumber
      & \hphantom{=}
      + 4\pi \alpha\lp( \rho_E + S \rp),
\\
\label{em_maxwell_lm_gh_bssn}
      \lm \gh_{ij} &= -2 \alpha \Ah_{ij} - \frac{2}{3} \gh_{ij} 
      \dh_{m}\beta^{m},
\\
\label{em_maxwell_lm_ah_bssn}
      \lm \Ah_{ij} &= \e4c \lp[-D_i D_j\alpha  + \alpha R_{ij} 
      -8\pi\alpha S_{ij}\rp]^{\mathrm{TF}} 
      \\ \nonumber 
      & \hphantom{=}
      - \frac{2}{3} \Ah_{ij} \dh_m \beta^m + \alpha\lp( K \Ah_{ij} 
      - 2\Ah_{ik} \Ah^{k}\vp_{j} \rp),
\\
\label{em_maxwell_lm_Dh_bssn}
      \lm \Lh^i &= \gh^{mn} \Do_m \Do_n \beta^i -2\Ah^{im}\dh_m\alpha
      \\ \nonumber
      &\hphantom{=}
      +2\alpha\lp( 6 \Ah^{ij} \dh_j \chi - \frac{2}{3} \gh^{ij}\dh_j 
      K - 8 \pi \jh^i \rp)
      \\ \nonumber
      &\hphantom{=}
      + \frac{1}{3} \lp[ \dh^i\lp(\dh_n \beta^n\rp) + 2\Lh^i \dh_n\beta^n\rp]
      \\ \nonumber
      &\hphantom{=}
      +2 \alpha \Ah^{mn} \Dh^{i}\vp_{mn}.
\end{align}

These equations introduce two additional metrics: the conformal metric 
$\gh_{ij}$,
\begin{align}
   \gh_{ij} &= e^{-4\chi}\g_{ij} 
   =
   \begin{pmatrix}
   g_a       & \rho g_d  & 0          \\
   \rho g_d  & g_b       & 0          \\
   0         & 0         & \rho^2 g_c 
   \end{pmatrix},
\end{align}
and a flat reference metric $\go_{ij}$,
\begin{align}
   \go_{ij} &= 
   \begin{pmatrix}
   1       & 0     & 0          \\
   0       & 1     & 0          \\
   0       & 0     & \rho^2 
   \end{pmatrix},
\end{align}
which shares the same divergence characteristics as $\gh_{ij}$ and serves to 
regularize several quantities related to the contracted Christoffel symbols. 
In (\ref{em_maxwell_lm_chi_bssn})--(\ref{em_maxwell_lm_Dh_bssn}), hats denote 
quantities raised with $\gh^{ij}$ while $\dh$ and $\Do$ denote covariant 
differentiation with respect to the conformal metric and flat reference metric,
respectively.

In (\ref{em_maxwell_lm_ah_bssn}), the Ricci tensor is split into conformal 
and scale components via
\begin{align}
   R_{ij} &= \Rh_{ij} + R^{\chi}_{ij},
\\
   \Rh_{ij} &= -\frac{1}{2}\gh^{mn}\Do_m\Do_n\gh_{ij} + \gh_{m(i}\Do_{j)}
   \Lh^m 
   \\ \nonumber
   &\hphantom{=}
   + \Lh^m\Dh_{(ij)m} +2\Dh^{mn}\vp_{(i}\Dh_{j)mn} 
   \\ \nonumber
   &\hphantom{=}
   + \Dh^{mn}\vp_{i}\Dh_{mnj},
\\
   R^{\chi}_{ij} &= -2 \dh_i\dh_j \chi - 2\gh_{ij}\dh^{k}\dh_k \chi 
   + 4\dh_i\dh_j \chi 
   \\ \nonumber
   &\hphantom{=}
   - 4 \gh_{ij} \dh^k \chi\dh_k \chi \, .
\end{align}

We note that in an appropriate gauge, the GBSSN variables have no unstable 
growing modes associated with constraint violation 
\cite{mongwane2016hyperbolicity, cao2022note}. The Hamiltonian, momentum and 
contracted Christoffel constraints take the form
\begin{align}
   H &= \frac{1}{2} \lp( R + \frac{2}{3} K^2 - \Ah_{ij}\Ah^{ij} \rp)
   - 8 \pi \rho,
   \\
   M^i &= \e4c\lp( \dh_j \Ah^{ij} - \frac{2}{3} \gh^{ij}\dh_j K 
   + 6 \Ah^{ij}\dh_{j} \chi 
   \rp.
   \\ \nonumber
   &\hphantom{=}\lp.
   - 8\pi \jh^i \rp) =\lp(M^z, \rho M^\rho, 0\rp),
   \\
   {Z}^i &= \Lh^{i} - \Dh^{i} = \lp({Z}^z, 
   \rho {Z}^\rho, 0\rp).
\end{align}
It is worth noting that in 
(\ref{em_maxwell_lm_chi_bssn})--(\ref{em_maxwell_lm_Dh_bssn}) we have not 
included the usual dimensionful constraint damping parameters. The critical 
solutions we investigate have no single length scale and our code must be 
able to deal with solutions spanning many orders of magnitude in scale.
By choosing a set of damping parameters which worked well at a given scale, we 
might have introduced inconsistent and difficult to debug behaviours at other 
scales. These might include: 
\begin{itemize}
\item Improved constraint conservation in the long wavelength regime at the 
expense of the short wavelength regime.
\item Unexpected interactions with Kreiss-Oliger dissipation 
\cite{kreiss1973methods}.
\item Scale dependent issues arising at grid boundaries due to suboptimally 
chosen adaptive mesh refinement (AMR) parameters.
\end{itemize}
In order to avoid these possibilities and to ensure our that 
code had no preferential length scale, we omitted the damping parameters in 
our simulations.

In summary, the complete set of geometric variables is given by the lapse, 
$\alpha$, shift, $\beta^{i}$,
\begin{align}
   n^\mu &= \lp(\frac{1}{\alpha}, -\frac{\beta^{i}}{\alpha} \rp),
\\
   \beta^{i}&=\lp(\beta^z, \rho \beta^\rho, 0\rp),
\end{align}
conformal factor, $\chi$, conformal metric, $\gh_{ij}$,
\begin{align}
   \gh_{ij} &= e^{-4\chi}\g_{ij} 
   =
   \begin{pmatrix}
   g_a       & \rho g_d  & 0          \\
   \rho g_d  & g_b       & 0          \\
   0         & 0         & \rho^2 g_c 
   \end{pmatrix},
\end{align}
trace of the extrinsic curvature, $K$, conformal trace-free extrinsic 
curvature, $\Ah_{ij}$,
\begin{align}
   \Ah_{ij} &= e^{-4\chi}\lp( K_{ij} - \frac{1}{3} \g_{ij} K \rp) ,
   \\
   &=
   \begin{pmatrix}
   A_a       & \rho A_d  & 0          \\
   \rho A_d  & A_b       & 0          \\
   0         & 0         & \rho^2 A_c 
   \end{pmatrix},
\end{align}
the quantities $\Delta^{i}$ representing the difference between the 
contracted Christoffel symbols of the conformal metric (${\Gh^{i}}_{jk}$)  
and flat reference metric (${\Go^{i}}_{jk}$),
\begin{align}
   {\Dh^{i}}_{ij} &= {\Gh^{i}}_{jk} - {\Go^{i}_{jk}},
   \\
   \Dh^{i} &= \Gh^i - {\Go^{i}}_{jk}\gh^{jk},
   \\ 
   \Dh^{i}&=\lp(\Dh^z, \rho \Dh^\rho, 0\rp),
\end{align}
and finally, the quantities $\Lambda^i$, representing the quantities 
$\Delta^i$ promoted to independent dynamical degrees of freedom rather than 
being viewed as functions of $\gh_{ij}$ and $\go_{ij}$,
\begin{align}
   \Lh^{i}&=\lp(\Lh^z, \rho \Lh^\rho, 0\rp).
\end{align}
Here, as is the case for the spacetime 4-metric, all of the GBSSN functions 
are taken to have even character about $\rho = 0$. For a more in-depth 
review of the GBSSN formulation we refer the reader to the works of Brown 
\cite{brown2009covariant} and Alcubierre et 
al.~\cite{alcubierre2011formulations}.

In our investigations of critical behaviour we consider both the massless 
scalar field and the Maxwell field. In the first instance, we have the 
Einstein equations and stress tensor,
\begin{align}
   G_{\mu \nu} &= 8\pi T^{\rm S}_{\mu \nu},
\\
\label{em_wave_stress_energy}
   T^{\rm S}_{\alpha \beta} &= \n_{\alpha}\mu\n_{\beta}\mu - \frac{1}{2} 
   g_{\alpha \beta} \n_{\gamma}\mu \n^{\gamma}\mu,
\end{align}
and a matter equation of motion,
\begin{align}
\label{em_wave_equation}
   \n^{\alpha}\n_{\alpha}\mu &= 0 \, .
\end{align}
For the Einstein-Maxwell system we have
\begin{align}
   G_{\mu \nu} &= 8\pi T^{\rm EM}_{\mu\nu},
\\
   \label{em_maxwell_Tuv}
   T^{\rm EM}_{\mu\nu} &= {F_{\mu}}^{\alpha}F_{\nu \alpha} - \frac{1}{4}g_{\mu\nu}
   F_{\alpha\beta}F^{\alpha\beta} ,
\end{align}
and matter equations of motion,
\begin{align}
   \label{em_maxwell_basic}
   \n_\mu F^{\nu\mu} &= 0,
\\
   \label{em_maxwell_dual_basic}
   \n_\mu \sF^{\nu\mu} &= 0.
\end{align}
Here,
\begin{align}
   \sF^{\mu\nu} &= \frac{1}{2} \eps^{\mu\nu \gamma\delta}F_{\gamma\delta},
\end{align}
with
\begin{align}
   \eps^{\alpha \beta \gamma \delta} &= \frac{\lp(-1\rp)^{s}}{\sqrt{|g|}}
   \epsilon^{\alpha \beta \gamma \delta} ,
\end{align}
where $s=1$ is the metric signature and $\epsilon^{\alpha \beta \gamma 
\delta}$ is the 4D Levi-Civita symbol.

Rather than use (\ref{em_maxwell_basic})--(\ref{em_maxwell_dual_basic}) and a 
vector potential decomposition of $F_{\mu\nu}$, we incorporate the 
source-free Maxwell equations into a larger system, similarly to how the GBSSN 
and FCCZ4 formalisms embed general relativity within variations of the Z4 
system~\cite{alcubierre2011formulations, daverio2018apples}. In the case of 
general relativity, this embedding enables the Hamiltonian and momentum 
constraints to be expressed through propagating degrees of freedom. 
Analogously, for the Maxwell fields, the divergence conditions become tied 
to propagating degrees of 
freedom~\cite{palenzuela_curved_em_2010, komissarov_em_2007}:
\begin{align}
   \label{em_maxwell_mod}
   -\sigma n^\nu\Psi_E &= \n_\mu \lp( F^{\nu\mu} +g^{\nu\mu}\Psi_E \rp),
\\
   \label{em_maxwell_dual_mod}
   -\sigma n^\nu\Psi_B &= \n_\mu \lp( \sF^{\nu\mu} + g^{\nu\mu}\Psi_B \rp).
\end{align} 
Here, $\sigma$ is a dimensionful damping parameter and $\Psi_{E}$ 
and $\Psi_{B}$ are constraint fields which couple to the violation 
of the divergence conditions for the electric and magnetic fields,
respectively. By promoting the constraints to propagating degrees of freedom, 
our solutions gain additional stability and exhibit advection and damping of 
constraint violations which would otherwise  accumulate. Finally, we take the 
following definitions of the electric fields, 
$E^{\alpha}$, magnetic fields, $B^{\alpha}$, and Maxwell tensors, 
$F^{\alpha \beta}$ and $\sF^{\alpha\beta}$,
\begin{align}
   E^\alpha &= F^{\alpha \beta}n_\beta,
\\
   B^\alpha &= \sF^{\beta \alpha}n_\beta,
\\
   F^{\alpha \beta} &= n^{\alpha}E^{\beta} - n^{\beta}E^{\alpha}
   +\eps^{\gamma \delta \alpha \beta}n_{\gamma}B_{\delta}
\\
   \sF^{\alpha \beta} &= n^{\beta}B^{\alpha} -n^{\alpha}B^{\beta} 
   +\eps^{\gamma \delta \alpha \beta}n_{\gamma}E_{\delta}
\end{align}
 
The evolution equations for the electric and magnetic fields, and the 
constraint variables, now take the form:
\begin{align}
   \lm E^i &= \eps^{ijk} D_j \lp( \alpha B_k \rp) + \alpha K E^i 
   + \alpha \g^{ij} D_j \Psi_E,
\\
   \lm B^i &= -\eps^{ijk} D_j \lp( \alpha E_k \rp) + \alpha K B^i 
   - \alpha \g^{ij} D_j \Psi_B,
\\
   \lm \Psi_E &= \alpha D_i E^i,
\\
   \lm \Psi_B &= -\alpha D_i B^i,
\end{align} 
where we have once again set dimensionful damping parameters to zero to avoid 
setting a preferential length scale. Under the restriction to  axisymmetry, 
the electric, magnetic, and associated fields simplify as
\begin{align}
   E^i & = \lp( 0, 0, E^\phi \rp),
\\
   B^i & = \lp(B^z, \rho B^\rho, 0\rp),
\\ 
   \Psi_{E} &=0,
\\
   \sF_{\mu\nu}F^{\mu\nu} &= -F_{\mu\nu}F^{\mu\nu}.
\end{align}
 
Similarly to the GBSSN functions, all of $B^z$, $B^\rho$, $E^\phi$, $\Psi_{B}$ 
and $F_{\mu\nu}F^{\mu\nu}$ are constructed to be even about the $\rho = 0$ 
axis. As is the case for the Hamiltonian, momentum and contracted Christoffel 
constraints of GBSSN, $\Psi_{B}$ and $\Psi_{E}$ evolve stably and vanish in the 
continuum limit provided the initial data obeys the relevant constraints.  

\section{\label{em_sec_initial_data}Initial Data}

We assume time symmetry on the initial slice such that $K_{ij} = j^{i} = 0$ 
with the momentum constraints automatically satisfied. Thus, our initial data 
represents a superposition of ingoing and outgoing solutions of equal 
magnitude and implies the existence of a family of privileged, on-axis, 
inertial observers who are likewise stationary at the initial time. Through 
careful construction, the geodesics these observers follow enable us to 
extract information concerning the evolution of our critical systems in a 
way that is completely independent of gauge.

Under the York-Lichnerowicz conformal decomposition and given time symmetry, 
the $t=0$ Hamiltonian constraint takes the form
\begin{align}
   \label{em_maxwell_gen_id_id_poisson}
   2H&=8\dh_i\dh^i e^{\chi} -\Rh e^{\chi}+ 16 \pi e^{5\chi} \rho = 0.
\end{align}

We choose the initial conformal 3-metric to be flat and isotropic and 
define the electric and magnetic fields as $E^i =e^{-9\chi/2} 
\tilde{E}^{i}$, $B^i = e^{-9\chi/2} \tilde{B}^{i}$  with $\tilde{E}^i$ and 
$\tilde{B}^i$ specified according to some initial profiles. These choices 
greatly simplify~(\ref{em_maxwell_gen_id_id_poisson}), and upon defining
\begin{align}
   \kappa &= e^{\chi},
\\
   \Do^2 \kappa &=\lp(\p_{\rho\rho} \kappa + \frac{\p_\rho \kappa}{\rho} 
   + \p_{zz} \kappa\rp),
\end{align}
the elliptic equation for the Einstein-Maxwell system takes the form
\begin{align}
\label{em_maxwell_id_kappa_maxwell}
   \Do^2 \kappa &= -\pi \lp( \rho^2 {\tilde{B}^{\rho}}^2 
   + {\tilde{B}^{z}}^2  + \rho^2{\tilde{E}^\phi}^2 \rp),
\end{align}
while the corresponding equation for the scalar field is
\begin{align}
\label{em_maxwell_id_kappa_scalar}
  \Do^2\kappa  &= - \pi \kappa \lp( \lp(\p_z\mu\rp)^2 
   + \lp(\p_\rho\mu\rp)^2  \rp).
\end{align}

In the case of the massless scalar field or pure electric field we are free 
to simply specify $\mu$ or $\tilde{E}^\phi$. For the case of a pure magnetic 
field we must additionally satisfy $D_i B^i = 0$. 
Under the transform $\bar{B}^i = e^{-6\chi} B^{i} = e^{-3/2 \cdot \chi}
\tilde{B}^i$, $B$ may be trivially derived from a vector potential via 
$\bar{B} = -\no \times A$.  Taking $A^z = 0$ and $A^\rho = 0$ results in 
families which satisfy the relevant constraints. Initially stationary magnetic 
type data is then specified via
\begin{align}
\label{em_vector_pot_def}
   A^i &= \lp(0,0,A^{\phi} \rp),
\\
\label{em_vector_pot_Bz_def}
   \bar{B}^z &= -\p_{\rho}{A^\phi} - \frac{A^\phi}{\rho},
\\
\label{em_vector_pot_Bp_def}
   \bar{B}^\rho &= \frac{1}{\rho} \p_z{A^\phi}.
\end{align} 

The initial data for the collapse of the massless scalar field is given in 
Table~\ref{em_wave_initial_data_families}. We make use of the function
\begin{align}
   g\lp(z,\rho, \rho_0, \lambda\rp) &= \exp\lp({-\frac{z^2+\lp(\rho 
   + \rho_0\rp)^2}{\lambda^2}}\rp),
\end{align}
and present all initial data in a manner which is manifestly scale 
invariant with respect to the parameter $\lambda$: under a 
rescaling $\lambda \rightarrow \lambda/\lambda'$ all dimensionless 
quantities ${f(t,z,\rho)}$ transform as ${f(t,z,\rho)} 
\rightarrow {f(t/\lambda', z/\lambda', \rho/\lambda')}$.

\begin{table}[!htb]
\centering
\def\arraystretch{1.5}
\begin{ruledtabular}
\begin{tabular}{ c  c  c }
    Family & Initial Data & $p^{\star}$
    \\ \hline
    $W_{l=0}$ 
    &
    $\mu = p \cdot \lp( g\lp(z,\rho, \rho_0, \lambda\rp) 
    + g\lp(z,\rho, -\rho_0, \lambda\rp)\rp)$
    &
    $\sim0.152$
    \\
    $W_{l=1}$ 
    &
    $\mu = p\cdot\frac{z}{\lambda} \lp( g\lp(z,\rho, \rho_0, \lambda\rp) 
    + g\lp(z,\rho, -\rho_0, \lambda\rp)\rp)$
    & 
    $\sim0.297$
\end{tabular}
\end{ruledtabular}
\caption{Families of initial data for  the massless scalar field. The form of 
the initial data is scale invariant with respect to $\lambda$ and we adopt 
$\lambda = 1$, $\rho_0 =0$ for all simulations. We refer to family $W_{l=0}$ 
as monopole type initial data and family $W_{l=1}$ as dipole type. The final 
column gives the approximate value of the critical parameter $p^{\star}$ for 
each family. 
}
\label{em_wave_initial_data_families}
\end{table}

For the Einstein-Maxwell system we investigate the families of initial data 
presented in Tables~\ref{em_maxwell_initial_data_families} and 
\ref{em_paper_initial_data_families}.  The families given in 
Table~\ref{em_maxwell_initial_data_families} are new to this work
while those given in Table~\ref{em_paper_initial_data_families} 
correspond to the dipole and quadrupole families 
of~\cite{mendoza_critical_phenomena_electromagnetic_2021}.
The families of Table~\ref{em_maxwell_initial_data_families} were chosen in the 
hope that that similarities and differences in the underlying behaviors of 
dipole ($l=1$) and quadrupole ($l=2$) solutions would reveal information 
concerning the universality of the critical solutions. The two families of 
dipole initial data ($E_{l=1}$ and $M_{l=1}$) correspond to electric and 
magnetic dipoles, respectively, and are initially quite dissimilar. 

\begin{table}[!htb]
\centering
\def\arraystretch{1.5}
\begin{ruledtabular}
\begin{tabular}{ c  c c }
    Family & {Initial Data} & $p^{\star}$
    \\ \hline
    $E_{l=1}$
    & $\tilde{E}^\phi = p\cdot\frac{1}{\lambda^2} 
    \lp(g\lp(z,\rho, \rho_0, \lambda\rp) 
    + g\lp(z,\rho, -\rho_0, \lambda\rp)\rp)$
    & $\sim0.644$ 
    \\
    $M_{l=1}$
    & ${A^\phi} =  p\cdot\frac{\rho}{\lambda} 
    \lp(g\lp(z,\rho, \rho_0, \lambda\rp) 
    + g\lp(z,\rho, -\rho_0, \lambda\rp)\rp)$
    & $\sim0.377$
    \\
    $M_{l=2}$
    & ${A^\phi} = p\cdot\frac{z\rho}{\lambda^2} 
    \lp(g\lp(z,\rho, \rho_0, \lambda\rp) 
    + g\lp(z,\rho, -\rho_0, \lambda\rp)\rp)$
    & $\sim0.896$
\end{tabular}
\end{ruledtabular}
\caption{Families of initial data for the Einstein-Maxwell system. The form 
of the initial data is scale invariant with respect to $\lambda$ and we refer 
to family $E_{l=1}$ as the electric dipole type, $M_{l=1}$ as the magnetic 
dipole type and $M_{l=2}$ as the magnetic quadrupole type. $\tilde{B}^i$ is 
determined from $A^{\phi}$ via 
(\ref{em_vector_pot_def})--(\ref{em_vector_pot_Bp_def}). 
All of our investigations adopt $\lambda = 1$ and $\rho_0  = 0$. 
We note that although $\tilde{E}^i$ and $A^i$ are pure multipoles, the 
initial spacetime is far from flat and, in fact, the evolution is initially 
in the strong-field regime.}
\label{em_maxwell_initial_data_families}
\end{table}

As stated,
the families of Table~\ref{em_paper_initial_data_families} correspond to those 
in~\cite{mendoza_critical_phenomena_electromagnetic_2021} where the initial 
data was presented in an orthonormal coordinate basis. Here we present it in 
terms of the tensor quantities $\bar{E}^i = e^{-6\chi} E^{i}$. Notably, we do 
not find the same critical points for the data in 
Table~\ref{em_paper_initial_data_families} as was found 
in~\cite{mendoza_critical_phenomena_electromagnetic_2021}. Instead of 
$p^{\star}_{\mathrm{dipole}} \approx 0.913$ and $p^{\star}_{\mathrm{quad}} 
\approx 3.53$, we find $p^{\star}_{\mathrm{dipole}} \approx 0.258$ and 
$p^{\star}_{\mathrm{quad}} \approx 0.997$. In light of the results of 
Sec.~\ref{em_subsec_direct_comparison} and since the ratios among the two 
family parameters are essentially identical, we suspect that either our
initial data or that 
of~\cite{mendoza_critical_phenomena_electromagnetic_2021} 
was simply scaled by some unaccounted for factor. 

\begin{table}[!htb]
\centering
\def\arraystretch{1.5}
\begin{ruledtabular}
\begin{tabular}{ c  c  c }
    Family & {Initial Data} & $p^{\star}$
    \\ \hline
    $E_{\mathrm{dipole}}$
    & $\bar{E}^\phi = p\cdot\frac{8}{\lambda^2} \exp\lp({ -
    \frac{z^2+\rho^2}{\lambda^2}}\rp)$
    & $\sim 0.258$
    \\
    $E_{\mathrm{quad}}$
    & $\bar{E}^\phi = p\cdot\frac{16z}{3\lambda^3} \exp\lp({ -
    \frac{z^2+\rho^2}{\lambda^2}}\rp)$
    & $\sim 0.997$
\end{tabular}
\end{ruledtabular}
\caption{Families of initial data specified 
in~\cite{mendoza_critical_phenomena_electromagnetic_2021}. Here, we have 
expressed the initial data in standard tensor notation, rather than in an 
orthonormal basis as 
in~\cite{mendoza_critical_phenomena_electromagnetic_2021}, so that $p$ 
is a dimensionless strength parameter. }
\label{em_paper_initial_data_families}
\end{table}

\section{\label{em_sec_validation}Numerics and Validation}

We calculate the initial data using~(\ref{em_maxwell_id_kappa_maxwell}) 
and (\ref{em_maxwell_id_kappa_scalar}) with a multigrid method on a 
spatially compactified grid,
\begin{align}
   z &= \tan\lp(\frac{Z\pi}{2}\rp), \quad -1 \le Z \le 1 \, ,
\\
   \rho &= \tan\lp(\frac{R\pi}{2}\rp), \quad 0 \le R \le 1 \, ,
\end{align}
which renders the outer boundary conditions trivial. A consequence of this 
transform is that the eigenvalues of the finite difference approximations of 
(\ref{em_maxwell_id_kappa_maxwell}) and (\ref{em_maxwell_id_kappa_scalar}) 
become highly anisotropic: for a number of grid points that provides 
adequate accuracy, the characteristic magnitude of the action of the 
differential operator on the grid function $\kappa$ may be as much as 
$1\cdot10^7$ times larger near the edge of the grid as it is at the origin. 
To account for the resulting large eigenvalue anisotropy, line relaxation 
is employed to increase convergence rates. 

An unfortunate side effect of using a global line relaxation technique
is that at our chosen resolution, and for a tuning precision 
$\lp|p^\star-p|\rp/p^\star \lesssim 1\cdot10^{-9}$, we lose the ability 
to discriminate between sets of initial data. That is, the price we pay for 
global relaxation of the highly anisotropic problem is a loss of precision.
We resolve this issue by calculating three reference solutions corresponding to 
parameters $p_1$, $p_2$ and $p_3$ near the critical point, $p^\star$, such that 
$\lp|p^\star-p_i|\rp/p^\star \approx 1 \cdot10^{-6}$ and $\lp|p_j-p_i|\rp/
p^\star \approx 1 \cdot10^{-6}$ for $i \neq j$. For simulations with 
$\lp|p^\star-p|\rp/p^\star \lesssim 1 \cdot10^{-6}$, initial data is then 
calculated via third order point-wise spatial interpolation of grid 
functions using the three reference solutions. The error thereby introduced 
is orders of magnitude below that of the numerical truncation error in the 
subsequent evolution and may be safely ignored.

Our evolution code is built on a slightly modified version of 
PAMR~\cite{pamr_reference} and AMRD~\cite{amrd_reference}. We use a second 
order in space and time integrator with Kreiss-Oliger dissipation terms to 
damp high-frequency solution components. Additional resolution is allocated 
as required through the use of AMR based on local truncation error estimates.

Close to criticality, these simulations made heavy use of AMR. A run for 
family $M_{l=1}$, for example, would have a base resolution of $[129,129]$ 
with four levels of 2:1 refinement at $t=0$. At the closest approach to 
criticality ($\lp| p^\star-p\rp|/p^\star \approx 1\cdot10^{-13}$), the 
simulation would have $\sim20$ levels of refinement representing an increase 
in resolution on order of 10,000.  

The code was originally based on a fourth order in space and time 
method. During the course of our investigations we found that, without great 
care, spatial differentiation in the vicinity of grid boundaries could easily 
become pathological for higher order integration schemes and this was 
particularly true when we used explicit time integration. 
Without careful consideration, these sometimes subtle effects could completely 
negate any advantages gained from the use of a higher order scheme. As a 
result, the decision was made to employ a much easier to debug second order 
accurate method. Specifically, we opted to use a second order Runge-Kutta 
integrator with second order accurate centered spatial differencing and fourth 
order Kreiss-Oliger dissipation \cite{kreiss1973methods}. 
In order to reduce the effect of spurious reflection from AMR boundaries, we 
employ a technique very similar to that of Mongwane~\cite{mongwane2015toward}.

\subsection{\label{em_subsec_gauge}Choice of Gauge}

Our evolution code accommodates a wide variety of hyperbolic gauges with most 
of our investigations focusing on versions of the standard Delta driver and 
1+log families of shift and slicing conditions~\cite{Baumgarte_1998_mp_1, 
alcubierre3071_mp_2, campanelli2006_mp_4, alcubierre2003_mp_5}. We found that 
there were no significant issues associated with using various  Delta driver 
shifts for evolutions moderately close to criticality ($|p-p^\star|/p^\star 
< 1\cdot10^{-3}$), but that their use tended to significantly increase the 
grid resolution, and therefore computational cost, required to resolve the 
solutions. As such, the results presented in Secs.~\ref{em_subsec_convergence} 
and \ref{em_sec_results} are based on the following choice of gauge:
\begin{align}
   \lm \alpha &= -2 \alpha K ,
\\
   \beta^r &= 0,
\\
   \beta^z &= 0.
\end{align}

\subsection{\label{em_subsec_classification}Classification of Spacetimes}

We characterize spacetimes as either subcritical or supercritical based on 
two primary indicators: the dispersal of fields and the collapse of the 
lapse. While the more definitive approach to flagging a spacetime containing
a black hole would be to identify an apparent 
horizon, we opt for monitoring the lapse collapse due to its simplicity and 
practicality. One drawback of this approach is the potential ambiguity in the 
final stages of the last echo in each family. 
Specifically, it is unclear 
whether the behaviour that is observed for putatively marginally supercritical
collapse represents a genuine physical singularity or merely a coordinate 
artifact. However, by closely observing the growth trends of invariant 
quantities and confirming the dispersion of sub-critical solutions, we are 
confident that our results, up to the final portion of the last echo, depict 
the genuine approach to criticality. Given the inherent challenges in precisely 
determining $p^{\star}$, we have chosen to exclude the simulations closest to 
criticality when computing values for $\g$ and $\Delta$ across all families 
of initial data.

\subsection{\label{em_subsec_gbssn}GBSSN Considerations}

Aside from the standard convergence and independent residual convergence 
tests (Section~\ref{em_subsec_convergence}), it is important to quantify the 
behaviour we expect from a code based on the GBSSN formalism when in the 
critical regime. First and foremost, in their most general forms (without 
enforcing elliptic constraints) GBSSN evolution schemes are unconstrained. 
We should therefore expect constraint violations to grow with time while 
remaining bounded and convergent for well resolved initial data sufficiently 
far from criticality.

Another potentially overlooked factor concerning the GBSSN formulation is 
that GBSSN is not only over-determined (e.g.~the evolution equations for 
$\hat{\Lambda}^i$  are implicit in the evolution of the other variables and 
the maintenance of the constraints), but GBSSN is effectively an embedding 
of general relativity within a larger Z4 type system under the assumption 
that the Hamiltonian constraint holds~\cite{alcubierre2011formulations, 
daverio2018apples}. In practice, this means that a well resolved and convergent 
solution in GBSSN may cease being a valid solution within the context of 
general relativity at some point during the evolution. This is perhaps best 
illustrated by considering the near critical evolution of the Einstein-Maxwell 
system depicted in Fig.~\ref{em_num_meth_em_evo_cstr_1}. Although the 
Hamiltonian and momentum constraints are well maintained throughout the 
evolutions, the final ``dispersal'' state is not a valid solution in the 
context of general relativity. In this case, the constraint violations of the 
overdetermined system have made it so that the geometry that remains as the 
electromagnetic pulse disperses to infinity is a constraint-violating remnant 
rather than flat spacetime.

The overall effect is that our solutions cannot be trusted for particularly 
long periods of time after they make their closest approach to criticality.
This in turn presents obvious difficulties in determining the mass of any 
black holes in the supercritical regime where it may take significant 
coordinate time for the size of the apparent horizon to approach that of the 
event horizon. For this reason we restrict our analysis to the subcritical 
regime.   

To verify that we remain ``close'' to a physically meaningful general 
relativistic solution, we monitor the magnitude of the constraint violations 
relative to quantities with the same dimension. We also monitor independent
residuals for the fundamental dynamical variables. We consider a 
solution using AMR to be reasonably accurately resolved when:
\begin{itemize}
\item The independent residuals and constraints violations of an AMR solution 
in the strong field (nonlinear) regime are maintained at levels comparable 
to those determined from convergence tests using uniform grids.
\item The independent residuals are kept at acceptable levels relative to the 
magnitude of the fields.
\item The constraint violations are kept small relative to the magnitude of 
their constituent fields. (e.g. $|H| \ll |R|$). 
\end{itemize}  

For a dispersal solution close to the critical point, the second and third of 
these conditions are guaranteed to fail some period of time after the solution 
makes its closest approach to criticality. Thankfully, in practice we have 
found that with adequately strict truncation error requirements (relative 
truncation errors below $1\cdot 10^{-3}$ seems sufficient and we maintain 
$5\cdot10^{-5}$ for all simulations), the conditions remain satisfied 
throughout the collapse process.

\subsection{\label{em_subsec_convergence}Convergence}

The parameters for our convergence test simulations are given in 
Table~\ref{em_num_meth_em_conv_id_table}. Note that these simulations and 
those given in Section~\ref{em_sec_results} are performed on 
semi-compactified grids with
\begin{align}
   z = \sinh{Z'}, \quad 0 \le Z' \le 12 \, ,
\\
   \rho=\sinh{P'}, \quad 0 \le P' \le 12 \, .
\end{align}

For all of the calculations discussed in this paper, appropriate boundary 
conditions are set at $z=0$ to mirror or reflect the GBSSN and matter 
variables, depending on whether the given field has even or odd character 
about the $z=0$ plane. This simplification allows us to reduce 
the required compute time by a factor of two and alleviates issues which 
occasionally arise from asymmetric placement of AMR boundaries. For these and 
all subsequent results, initial data was calculated on a fully compactified 
grid as described in the introduction to this section.

\begin{table}[!h]
\centering
\begin{ruledtabular}
\begin{tabular}{ c  c  c  c  c  c  c  c  c }
    Family & Level & $p$ & $P_\mathrm{min}$ & $P_\mathrm{max}$ & $N_P$ 
    & $Z_\mathrm{min}$ & $Z_\mathrm{max}$ & $N_Z$\\
    \hline
    $M_{l=1}$ & 1 & 0.33 & 0 & 12 & 513  & 0 & 12 & 513 \\
    $M_{l=1}$ & 2 & 0.33 & 0 & 12 & 1025 & 0 & 12 & 1025 \\
    $M_{l=1}$ & 3 & 0.33 & 0 & 12 & 2049 & 0 & 12 & 2049 \\
    $M_{l=1}$ & 4 & 0.33 & 0 & 12 & 4097 & 0 & 12 & 4097
\end{tabular}
\end{ruledtabular}
\caption{ Parameters for magnetic dipole (family $M_{l=1}$) convergence 
tests. These simulations are well within the nonlinear regime with the 
critical point given by $p^\star \approx 0.377$. Similar convergence tests 
were performed for all families listed in 
Tables~\ref{em_wave_initial_data_families}--\ref{em_paper_initial_data_families}.
 }
\label{em_num_meth_em_conv_id_table}
\end{table}

Figs.~\ref{em_num_meth_conv_ham_psi_b}--\ref{em_num_meth_conv_mom-p-mom-z} 
demonstrate the convergence of the constraints for strong field dispersal 
solutions of the EM system. 
These figures additionally plot constraint violations for AMR simulations 
with a relative error tolerance of  $5\cdot10^{-5}$, demonstrating that the 
AMR simulations remain well within the convergent regime. The AMR 
simulations had an associated compute cost approximately 4 times larger than 
the lowest resolution unigrid simulations. 

Beyond monitoring the various 
constraints, we computed independent residuals of the various GBSSN 
quantities. The independent residuals were based on a second order in time 
and space stencil with three time levels and spatial derivatives evaluated 
at only the most advanced time. These residuals converged at second order 
as expected for all our tests.

Fig.~\ref{em_num_meth_em_evo_cstr_1} demonstrates the magnitude of 
various error metrics relative to the magnitude of the underlying fields. 
Throughout the collapse process, the solution is well resolved, but during 
dispersal ($t>6$), the solution becomes dominated by a non propagating
Hamiltonian constraint violation. Again, this is the expected behaviour for 
\mbox{GBSSN} type simulations where the Hamiltonian constraint is not tied to a 
dynamical variable or explicitly damped. In the limit of infinite 
resolution, we expect $R(t,0,0)$ and $H(t,0,0)$ to approach 0 at late times. 

We also note that, in addition to the GBSSN approach, we experimented with 
implementations of formulations derived from the Z4 formalism. 
In practice, we found that the use of Z4 formulations (without damping) 
resulted in significantly better constraint conservation post dispersal while 
exhibiting degraded Hamiltonian constraint conservation during collapse.
As we were predominantly interested in maintaining high accuracy during 
collapse, we opted to use {GBSSN} rather than, for example, fully covariant 
and conformal Z4 ({FCCZ4}).

Results similar to 
Figs.~\ref{em_num_meth_conv_ham_psi_b}--\ref{em_num_meth_em_evo_cstr_1} 
hold for all constraints and independent residual evaluators for each of the 
families $E_{l=1}$, $M_{l=1}$, $M_{l=2}$, $W_{l=1}$ and $W_{l=2}$. In all
cases convergence was second order as expected.

\begin{figure}[!ht]
\centering
\includegraphics[scale=1.0]{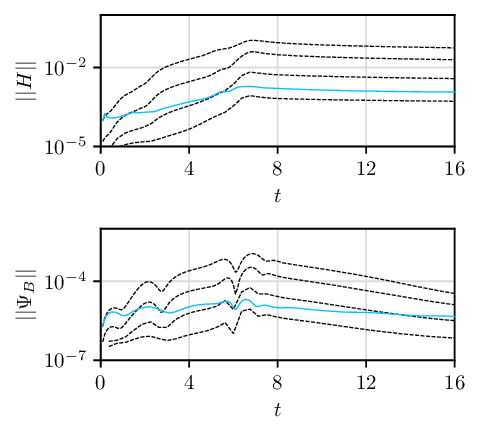}
\caption{Convergence of $l_2$ norms of $\Psi_{B}$ and  Hamiltonian constraint 
violations for strong field initial data given by 
Table~\ref{em_num_meth_em_conv_id_table}. The plotted norms of the residuals 
for each run are evaluated by interpolating the results to a uniform grid 
which has sufficient resolution to resolve the details of the simulation. 
This enables us to directly compare the convergence properties at the various 
resolutions. Each of the dashed lines represents a successive refinement (by 
a factor of 2) of the initial data while the solid line represents an AMR run 
with a relative error tolerance of $5\cdot10^{-5}$. The grid parameters for 
the various unigrid runs are given in 
Table~\ref{em_num_meth_em_conv_id_table}. }
\label{em_num_meth_conv_ham_psi_b}
\end{figure}

\begin{figure}[!ht]
\centering
\includegraphics[scale=1.0]{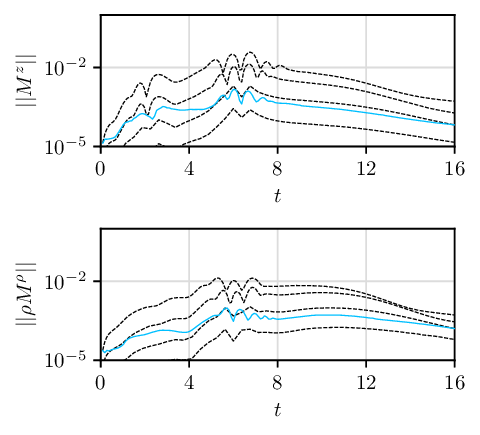}
\caption{As Fig.~\ref{em_num_meth_conv_ham_psi_b} but for the convergence
of the momentum constraints. Each of the dashed lines represents a successive 
refinement of the initial data while the solid line represents an AMR run 
with a relative error tolerance of $5\cdot10^{-5}$.  }
\label{em_num_meth_conv_mom-p-mom-z}
\end{figure}

\begin{figure}[!ht]
\centering
\includegraphics[scale=1.0]{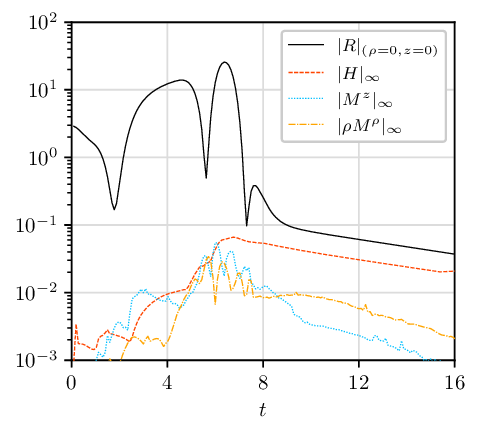}
\caption{ Magnitude of the 3D Ricci scalar, $R$, evaluated at $(0,0)$ 
and the $l_\infty$ norms of the Hamiltonian and momentum constraint 
violations for the AMR runs shown in 
Figs.~\ref{em_num_meth_conv_ham_psi_b} and \ref{em_num_meth_conv_mom-p-mom-z}.
Post-dispersal, the solution becomes dominated by a non dispersing 
Hamiltonian constraint violation. Our critical search AMR simulations maintain
constraint violations to about $1$ part in $500$ relative to the magnitude of
the relevant fields throughout collapse. }
\label{em_num_meth_em_evo_cstr_1}
\end{figure}   

\section{\label{em_sec_results}Results}

\subsection{\label{em_subsec_wave_results}Massless Scalar Field}

We choose to include simulations of massless scalar field collapse in order to 
test the accuracy of our simulations and to verify the utility of our analysis 
methods. Extremely high accuracy numerical analysis of $\Delta$ have 
determined that, for the case of spherically symmetric critical collapse, 
$\Delta \approx 3.4454524022278213500$~\cite{reiterer2019choptuik, 
matt_global_structure_choptuik} while $\gamma \approx 0.37$ is known from 
simulations~\cite{choptuik_universality_and_scaling_1993}. In this section, 
we verify that our simulations and analysis are of sufficient accuracy to 
reproduce these results. 

As specified in Table~\ref{em_wave_initial_data_families},
family $W_{l=0}$ is given by initially spherically symmetric initial data 
while family $W_{l=1}$ is initially a dipole. With family $W_{l=0}$, we 
demonstrate that our code is capable of resolving the spherically symmetric 
critical solution. By following the evolution of family $W_{l=1}$ we show 
that our code is capable of resolving situations where the initial data 
bifurcates into multiple on-axis centers of collapse. Since the results of 
Mendoza and Baumgarte~demonstrated that quadrupole initial data was 
subject to such a bifurcation, we felt that it was important to validate our 
code in a similar regime.
We have tuned these simulations to near the limits of double precision with 
$|p^{\star}-p|/p^{\star}\approx 1\cdot10^{-15}$ for family $W_{l=0}$ and 
$|p^{\star}-p|/p^{\star}\approx 1\cdot10^{-14}$ for family $W_{l=1}$.

Consider the proper time, $\tau$, of an inertial observer located at the 
accumulation point such that the observer would see the formation of 
a naked singularity at $\tau = \tau^{\star}$. The echoing period, $\Delta$, 
is then calculated using three somewhat independent methods. 
First, $\Delta_1$ is computed by taking the mean and standard deviation 
of the period between successive echoes at the center of collapse when viewed 
as a function of $-\ln\lp( \tau^\star - \tau \rp)$. Second, $\Delta_2$ comes 
from Fourier analysis of the dominant mode at the center of collapse in a 
similar frame. Third, $\Delta_3$ is calculated via the scaling 
relation~(\ref{em_dss_mass_scaling}), which results in an observer independent 
method given by
\begin{align}
   \label{em_results_Delta_scaling}
   \Delta_3&\approx\frac{\gamma}{N}\lp(\ln \lp| p_1-p^\star \rp| 
   -\ln \lp| p_2-p^\star \rp| \rp) \, .
\end{align}
Here, $N$ is the number of echoing periods observed between simulations 
with family parameters $p_1$ and $p_2$, respectively. 
Table~\ref{em_results_wave_results} summarizes the results using all three 
methods.

Plots of the lapse, $\alpha$, and the scalar field, $\mu$, as a function of 
logarithmic proper time  evaluated at the approximate accumulation points are 
shown in Figs.~\ref{em_results_wave_monopole} and \ref{em_results_wave_dipole} 
for families $W_{l=0}$ and $W_{l=1}$, respectively. Here, approximate 
accumulation points are defined as coordinate locations, $(z,\rho)$, of 
maximal scalar curvature encountered during the course of a subcritical 
simulation. These plots enable both direct and indirect calculation of 
$\Delta$ via the DSS time scaling relationship (\ref{em_dss_time_scaling}) 
and~(\ref{em_results_Delta_scaling}), respectively.

\begin{figure}[!htb]
\centering
\includegraphics[scale=1.0]{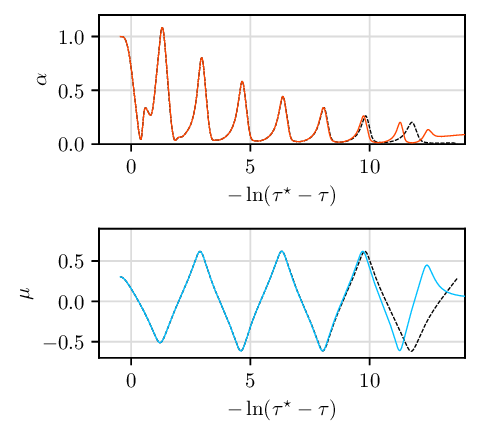}
\caption{ Lapse, $\alpha$, and value of the scalar field, $\mu$, as a 
function of $-\ln\lp( \tau^\star - \tau \rp)$ at the center of collapse 
(in this case the origin). Here, family $W_{l=0}$ data is used and both 
marginally subcritical (solid line) and supercritical (dashed line) solutions 
with  $\lp|{p^\star-p}\rp|/{p^\star} \approx 1\cdot10^{-15}$ are shown. 
Since the scalar field quickly approaches the critical solution with an 
associated strong-field scale that significantly decreases with each echo, 
we are able to accurately determine $\tau^\star$ to $\approx 1\cdot10^{-6}$. 
Direct measurement of $\Delta$ from $\mu$ gives  $\Delta_1 = 3.43(3)$, 
$\Delta_2 = 3.5(4)$ and  $\Delta_3 = 3.6(4)$. }
\label{em_results_wave_monopole}
\end{figure}   

\begin{figure}[!htb]
\centering
\includegraphics[scale=1.0]{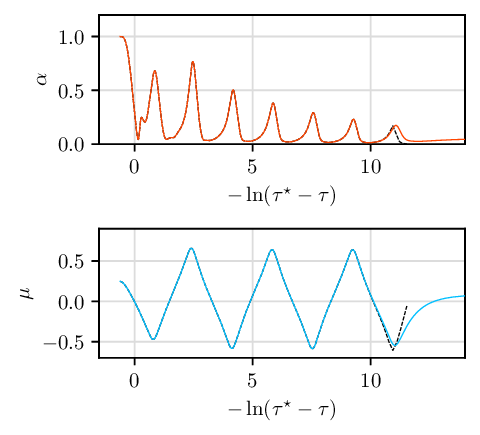}
\caption{ Lapse, $\alpha$, and value of the scalar field, $\mu$, as a function 
of $-\ln\lp( \tau^\star - \tau \rp)$ at the center of collapse (in this case 
$z\approx0.594$). In this case family $W_{l=1}$ data is used and the 
marginally subcritical (solid line) and supercritical (dashed line) solutions 
have been determined to an overall accuracy of $\lp|{p^\star-p}\rp|/{p^\star} 
\approx 1\cdot 10^{-14}$.  Here, $\tau^\star$ is computed to $\approx 
1\cdot10^{-5}$. Direct measurement of $\Delta$ from $\mu$ gives  
$\Delta_1 = 3.44(4)$, $\Delta_2 = 3.4(3)$ and $\Delta_3 = 3.2(4)$ }
\label{em_results_wave_dipole}
\end{figure} 

Unlike the $W_{l=0}$ case, we find that for the family $W_{l=1}$, the 
solution bifurcates into two centers of collapse. This in turn makes the 
determination of the world line of the privileged observer non-trivial. As 
we are starting from time symmetric initial data, the ideal solution would 
be to integrate the world lines of a family of initially stationary 
observers and choose the one which was nearest the accumulation point at the 
closest approach to criticality. Unfortunately, our code is not currently 
set up to perform such an integration. 

As a quick and potentially poor approximation, we choose the world line of an 
observer who remains at the approximate accumulation point throughout the 
evolution. This approximation is potentially error-prone because of its gauge 
dependence and the fact that the observer is generically non-inertial. 
However, for the case of the $W_{l=1}$ simulations, the solutions very quickly 
approaches two on-axis copies of the monopole solution so relatively little 
error appears to have been introduced by this choice.

The inverse Lyapunov exponent, $\gamma$, is calculated by fitting 
scaling laws of the form~(\ref{em_dss_rho_scaling}) for the maximal energy 
density, $\rho_{\mathrm{max}}$, and 3D Ricci scalar, $R_{\mathrm{max}}$, 
encountered during the course of a subcritical simulation. In these fits, we 
make the assumption that the dominant contribution to the putative universal 
periodic function is sinusoidal. As the specific region in parameter space 
where the scaling relationship is expected to hold is unknown (the 
uncertainty in $p^\star$ contaminates the values close to criticality, while 
radiation of dispersal modes contaminates the data far from criticality), 
we average a number of fits to reasonable subsets of the available data.

Ideally, we would calculate $\gamma$ via the maximal scale of some 
invariant quantity such as the 4D Ricci scalar (equivalently $\n_{\lambda} 
\mu \n^{\lambda} \mu$) or the Weyl scalar. However, calculations using frame 
dependent proxies such as the energy density, $\rho_E$, seem to be common in 
the literature and we have adopted this approach. In the case of collapse at 
the center of symmetry we note that $\rho_E$ is linearly related to the 
invariant $T$. Figs.~\ref{em_results_wave_gamma_rho} and 
\ref{em_results_wave_gamma_R} demonstrate the determination of $\gamma$ from 
$\rho_E$ and $R$ for families $W_{l=0}$ and $W_{l=1}$. Again, our results for 
$\gamma$ along with those for $\Delta_1$, $\Delta_2$ and $\Delta_3$ in the 
case of the scalar field are presented in Table~\ref{em_results_wave_results}.  
\begin{figure}[!ht]
\centering
\includegraphics[scale=1.0]{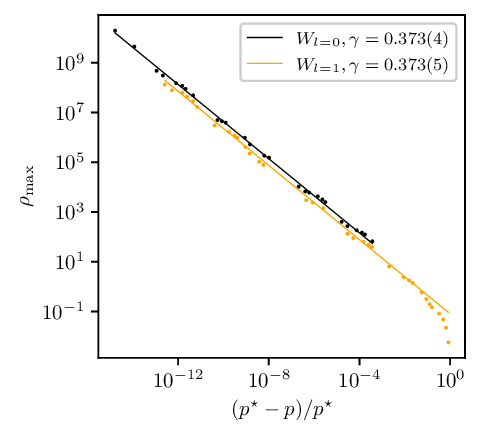}
\caption{ Inverse Lyapunov exponent, $\gamma$, determined via the scaling of 
the energy density $\rho_E$. Plotted here are the maximum values of $\rho_E$ 
obtained in each subcritical run as a function of 
$\lp|{p^\star-p}\rp|/{p^\star}$. The energy density has dimensions 
$M^{-2}$ and therefore scales according to $\lp| p^\star-p \rp|^{-2\gamma}$. 
Superior accuracy would be obtained by fitting to the maximum value of the 
4D Ricci scalar or another invariant quantity. The lines represent an 
averaged fit to the underlying data. }
\label{em_results_wave_gamma_rho}
\end{figure}    

\begin{figure}[!ht]
\centering
\includegraphics[scale=1.0]{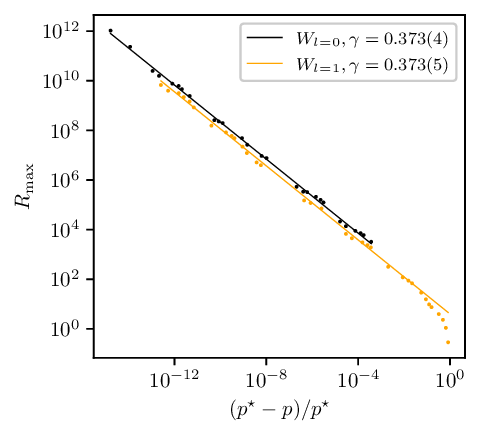}
\caption{ Inverse Lyapunov exponent, $\gamma$, determined via the scaling of 
the 3D Ricci scalar $R$. Plotted here are the maximum values of $R$ obtained 
in each subcritical run as a function of $\lp|{p^\star-p}\rp|/{p^\star}$. 
The Ricci scalar density has dimensions of $M^{-2}$ and therefore scales 
according to $\lp| p^\star-p \rp|^{-2\gamma}$. As in 
Fig.~\ref{em_results_wave_gamma_rho}, the lines represent an averaged fit 
to the underlying data. }
\label{em_results_wave_gamma_R}
\end{figure}    

\begin{table}[!ht]
\centering
\begin{ruledtabular}
\begin{tabular}{ c  c  c  c c }
    Family & $\Delta_1$ & $\Delta_2$ & $\Delta_3$ & $\gamma$
    \\
    \hline
    $W_{l=0}$  & $3.43(3)$ & $3.5(4)$  & $3.6(4)$ & $0.373(4)$
    \\
    $W_{l=1}$  & $3.44(4)$ & $3.4(3)$  & $3.2(4)$ & $0.373(5)$ 
    \\
\end{tabular}
\end{ruledtabular}
\caption{Estimated scaling exponents for axisymmetric scalar field collapse.
The results summarized here agree with previous investigations to within the 
estimated error of our calculations. Although $\Delta_3$ is far less precisely 
determined than $\Delta_1$, it can be found in the absence of knowledge 
concerning a privileged inertial observer.  }
\label{em_results_wave_results}
\end{table}

The excellent agreement between our computed values for the scaling exponents 
and previously established results for the massless scalar field demonstrate 
the accuracy of our simulations and the validity of our analysis. For AMR 
simulations, where it is impossible or impractical to establish the existence 
of convergence close to criticality, this process serves as an important 
verification and validation stage before the presentation of new results. 
It is worth noting that some previous studies \cite{choptuik2003_msf_axi, 
gundlach_critical_phenomena_2007, baumgarte2018aspherical} have presented 
evidence for a non-spherical unstable mode near criticality in scalar field
collapse.  
We see no evidence for such a mode for either our $W_{l=0}$ or 
$W_{l=1}$ calculations, but have not examined this point in much detail.

\subsection{\label{em_subsec_em_results}Einstein-Maxwell System}
With our methodology established and verified via investigation of the 
massless scalar field, the analysis of the critical collapse of the 
Einstein-Maxwell system proceeds in parallel fashion. We first consider the 
previously unstudied families $E_{l=1}$, $M_{l=1}$ and $M_{l=2}$ defined in 
Table \ref{em_maxwell_initial_data_families}. Once the behaviour of these 
solutions has been described, we turn our attention to the families of Table 
\ref{em_paper_initial_data_families} which were originally studied by 
Baumgarte et al.~\cite{baumgarte_critical_phenomena_electromagnetic_2019}, and 
Mendoza and Baumgarte~\cite{mendoza_critical_phenomena_electromagnetic_2021}. 
In what follows, we define the approximate accumulation points as the 
coordinate locations of maximal $|F_{\mu\nu} F^{\mu\nu}|$ encountered during 
a subcritical run. 

No bifurcations about the origin were observed for the dipole families 
$M_{l=1}$ and $E_{l=1}$: both families underwent collapse at the center 
of symmetry. Unfortunately, a gauge pathology prevented family 
$E_{l=1}$ from being investigated beyond $\lp|{p^\star-p}\rp|/{p}^{\star} 
\approx 5\cdot10^{-9}$. This shortcoming seems to bear some resemblance to 
the sort of gauge problems encountered in evolving Brill waves towards 
criticality~\cite{ledvinka_universality_of_curvature_invariants_2021} and may 
be able to be resolved through the use of the shock avoiding gauge 
suggested by Alcubierre in~\cite{alcubierre1997appearance} and successfully 
employed in~\cite{baumgarte2023critical, baumgarte2023criticalhexa}. 
Fortunately, the pathology occurs sufficiently late in the evolution to 
enable the extraction of meaningful information concerning $\Delta$ and 
$\gamma$ for the family. 

Figs.~\ref{em_results_em_m_dipole}--\ref{em_results_em_e_dipole} plot 
$\alpha$ and $|F_{\mu\nu}F^{\mu\nu}|$ at the accumulation point (in this 
case the origin) versus $-\ln{\lp(\tau^\star-\tau\rp)}$ for near-critical 
evolutions of families $M_{l=1}$ and $E_{l=1}$. Since the collapse occurs at 
the center of symmetry, there is only a single accumulation point and the 
observers at the origin are privileged and inertial. As mentioned previously, 
this enables $\Delta$ to be accurately determined via statistical and Fourier 
analysis.
\begin{figure}[!ht]
\centering
\includegraphics[scale=1.0]{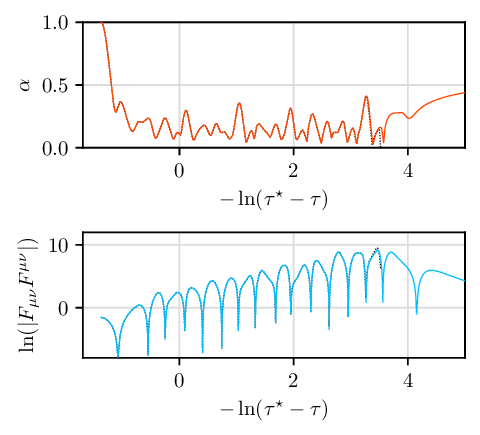}
\caption{ Lapse, $\alpha$, and invariant scalar, 
$\lp|F_{\mu\nu}F^{\mu\nu}\rp|$, at the center of collapse for family 
$M_{l=1}$ as a function of $-\ln\lp( \tau^\star - \tau \rp)$ for marginally 
subcritical (solid line) and supercritical (dashed line) solutions with 
$\lp| p^\star-p\rp|/p^\star \approx 1\cdot10^{-13}$. Unlike the case of the 
scalar field, the strong-field scale of the critical solution only slowly 
decreases (i.e.~$\Delta$ is small compared to the scalar case) and
$\tau^\star$ can only be determined to a relative tolerance of about 
$10^{-4}$. Direct measurement of $\Delta$ from $F_{\mu\nu}F^{\mu\nu}$ gives 
$\Delta_1 = 0.64(2)$ via statistical analysis, $\Delta_2 = 0.63(3)$ via 
Fourier analysis and $\Delta_3 = 0.59(6)$ 
from~(\ref{em_results_Delta_scaling}).  }
\label{em_results_em_m_dipole}
\end{figure}   

\begin{figure}[!ht]
\centering
\includegraphics[scale=1.0]{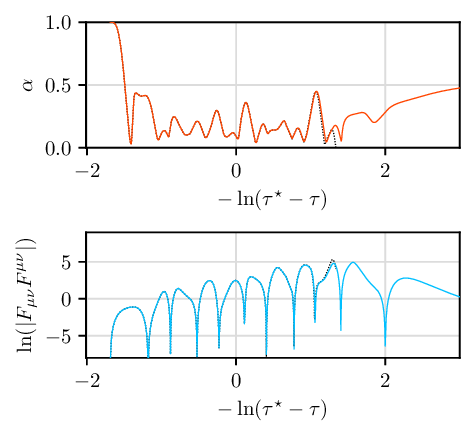} 
\caption{ Lapse, $\alpha$, and invariant scalar, 
$\lp|F_{\mu\nu}F^{\mu\nu}\rp|$, at the center of collapse for family 
$E_{l=1}$ as a function of $-\ln\lp( \tau^\star - \tau \rp)$ for marginally 
subcritical (solid line) and supercritical (dashed line) solutions with 
$\lp| p^\star-p\rp|/p^\star \approx 1\cdot10^{-9}$. As with family $M_{l=1}$,
the strong-field scale of the critical solution slowly decreases and 
$\tau^\star$ can only be determined to a relative tolerance of about 
$10^{-3}$. Direct measurement of $\Delta$ from $F_{\mu\nu}F^{\mu\nu}$ gives 
$\Delta_1 = 0.65(3)$ via statistical analysis, $\Delta_2 = 0.65(4)$ via 
Fourier analysis, and $\Delta_3 = 0.67(8)$ 
from~(\ref{em_results_Delta_scaling}). }
\label{em_results_em_e_dipole}
\end{figure}  

\begin{figure}[!ht]
\centering
\includegraphics[scale=1.0]{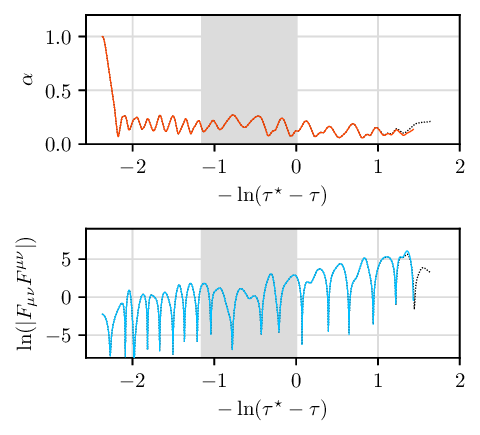}
\caption{ 
Lapse, $\alpha$, and invariant scalar, $\lp|F_{\mu\nu}F^{\mu\nu}\rp|$, at the 
center of collapse for family $M_{l=2}$ as a function of $-\ln\lp( \tau^\star 
- \tau \rp)$ with $\lp| p^\star-p\rp|/p^\star \approx 5\cdot10^{-13}$. We use 
the proper time of a gauge dependent accelerated observer located at 
$z \approx 0.440$ as our independent variable, $\tau$, making the 
interpretation of length and time scales potentially problematic. It appears 
that the critical solution is divided into two separate regions (transition 
region shown in gray) with differing $\Delta$ and $\gamma$. A naive 
measurement of $\Delta$ under the assumption that our observer is 
approximately inertial gives $\Delta_1 = 0.30(2)$ and $\Delta_2 = 0.31(3)$ for 
the first region and $\Delta_1 = 0.56(3)$, $\Delta_2 = 0.63(6)$ for the second 
region. Application of~(\ref{em_results_Delta_scaling}) (which is valid 
irrespective of the status of the observer) gives $\Delta_3 = 0.19(4)$ for the 
first region and $\Delta_3 = 0.64(9)$ for the second 
The values of $\Delta_2$ and $\Delta_3$ measured in the second region as 
$p \rightarrow p^{\star}$ appear to be consistent with those found for 
families $E_{l=1}$ and $M_{l=1}$.
}
\label{em_results_em_m_quad}
\end{figure} 

The analysis of family $M_{l=2}$ is both more interesting and more involved 
than that of families $M_{l=1}$ and $E_{l=1}$. In this case, and similarly to 
what is observed in the case of the massless scalar dipole, as the critical 
parameter is approached, the solution  bifurcates into two on-axis centers of 
collapse. After this bifurcation the character of the critical solution 
changes markedly. Specifically, following this transition period, the growth 
and echoing period of the separated collapsing regions come to resemble 
those of two separated copies of the $M_{l=1}$ or $E_{l=1}$ critical solutions. 
This change in character is somewhat obscured in time series plots by the 
fact that we use the proper time of an accelerated observer at the accumulation 
point rather than that of a privileged inertial observer. Despite this, the 
change is evident in the growth rate, $\gamma$, when calculated via the scaling 
relationship
\begin{align}
   \label{em_dss_FF_scaling}
   \ln{\lp(|F_{\mu\nu}F^{\mu\nu}|_{\rm max}\rp)} &= 
   -2\gamma\ln\lp|p-p^\star\rp| 
   \\ \nonumber
   & \hphantom{=}
   + f_{F}\lp(\gamma\ln\lp|p-
   p^\star\rp|\rp) + c_{F},
\end{align}
as well as when $\Delta$ is calculated via~(\ref{em_results_Delta_scaling}). 
Overall, the two distinct phases of collapse can be seen in 
Figs.~\ref{em_results_em_m_quad} and \ref{em_results_em_gamma}.

\begin{figure}[!ht]
\centering
\includegraphics[scale=1.0]{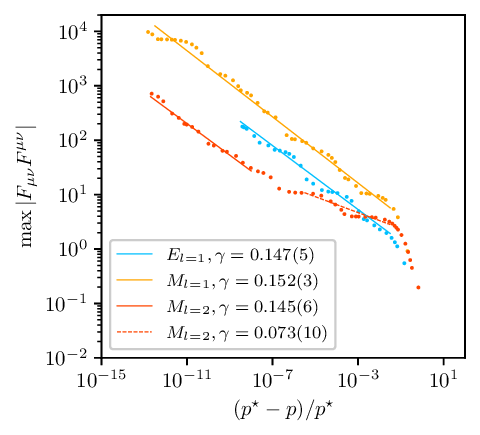}
\caption{ Inverse Lyapunov exponent, $\gamma$, determined via the scaling of
the invariant scalar $\lp|F_{\mu\nu}F^{\mu\nu}\rp|$ which should scale as 
$\lp| p^\star-p \rp|^{-2\gamma}$. Plotted here are the maximum values of 
the invariant obtained in each subcritical run as a function of 
$\lp|{p^\star-p}\rp|/{p^\star}$. 
The lines shown are averaged fits of the underlying data and the quoted 
values of $\gamma$ are the slopes of those fits. As described in the text, 
fits to two distinct regions of family $M_{l=2}$ have been made.  }
\label{em_results_em_gamma}
\end{figure}  

Figure~\ref{em_results_em_gamma} shows the results of calculating $\gamma$ via 
the scalar invariant $F_{\mu\nu}F^{\mu\nu}$ which should scale as 
$\lp| p^\star-p \rp|^{-2\gamma}$ as in (\ref{em_dss_FF_scaling}). Again, 
family $M_{l=2}$ appears to exhibit two distinct growth rates separated by a 
transition region in $\ln\lp|p^\star-p\rp|$. The early behaviour may be due 
to a slower growing quadrupole mode or perhaps simple radiation of initial 
data before the critical solution is approached. In total, the behaviour we 
observe appears to be consistent with the interpretation that, after the 
bifurcation occurs, the critical solution becomes dominated by the same mode 
as for families $M_{l=1}$ and $E_{l=1}$. A summary of our estimated values of 
$\Delta$ and $\gamma$ for the families defined in 
Table~\ref{em_maxwell_initial_data_families} is compiled in 
Table~\ref{em_results_em_results}.

\begin{table}[!ht]
\centering
\begin{ruledtabular}
\begin{tabular}{ c  c  c  c c }
    Family & $\Delta_1$ & $\Delta_2$ & $\Delta_3$ & $\gamma$
    \\
    \hline
    $E_{l=1}$  & $0.65(3)$ &  $ 0.65(4) $ 
    & $0.67(8) $ & $0.147(5)$
    \\
    $M_{l=1}$  & $0.64(2)$ &  $ 0.63(3) $ 
    & $0.59(6)$ & $0.152(3)$
    \\
    $M_{l=2}$ & $\mathbf{0.30(2)}$ & $ \mathbf{0.31(3)}$ 
    & $0.19(4)$ & $0.073(10)$
    \\
    $M_{l=2}$ & $\mathbf{0.56(3)}$ & $ \mathbf{0.63(6)}$ 
    & $0.64(9)$ & $0.145(6)$
    \\
    $E_{\mathrm{quad}}$  & $\bf{0.30(5)}$ &  $ \bf{0.33(2)} $ 
    & $0.61(11)
     $ & $0.164(19)$
    \\
    $E_{\mathrm{quad}}$  & $\bf{0.59(4)}$ &  $ \bf{0.57(4)} $ 
    & $0.59(10)
     $ & $0.152(20)$
\end{tabular}
\end{ruledtabular}
\caption{ Summary of computed scaling exponents in critical collapse of 
the EM field for the families presented in 
Tables~\ref{em_maxwell_initial_data_families} and 
\ref{em_paper_initial_data_families}. The analysis of $E_{\mathrm{quad}}$ 
is presented in Section \ref{em_subsec_direct_comparison}. Here, the two 
separate rows for $M_{l=2}$ and $E_{\mathrm{quad}}$  denote fits to the 
distinct behavioral regions of the quadrupole solutions; the first row 
is for $p$ fairly distant from $p^{\star}$ while the second is for 
$p \rightarrow p^{\star}$. Results in bold indicate that the measurements 
were made using the world line of an accelerated observer and are unlikely 
to be accurate. }
\label{em_results_em_results}
\end{table}

\subsection{\label{em_subsec_direct_comparison}Direct Comparison to Previous 
Work.}

When we compare our dipole and quadrupole results to those of 
Mendoza and Baumgarte~\cite{mendoza_critical_phenomena_electromagnetic_2021} 
and Baumgarte et 
al.~\cite{baumgarte_critical_phenomena_electromagnetic_2019}, the results are 
broadly consistent but do not fully agree to within our approximately 
determined errors. Although our work and the previous studies both indicate a
single unstable mode with $\gamma_{l=1} \approx 0.15$ and $\Delta_{l=1} 
\approx 0.6$ for dipole type initial data, our investigation into an 
alternative family of quadrupole type initial data is consistent with a 
universal (rather than family dependent) growth rate and echoing period. 
In order to more conclusively determine the consistency of our work with 
that of~\cite{mendoza_critical_phenomena_electromagnetic_2021} 
and~\cite{baumgarte_critical_phenomena_electromagnetic_2019}, we 
attempt to replicate the previous computations by performing critical 
searches for the families listed in Table~\ref{em_paper_initial_data_families}.

We perform evolutions of $E_{\mathrm{quad}}$ to a tolerance of $\lesssim4
\cdot10^{-15}$ so as to resolve the critical solution as accurately as 
possible. Previously, this family was resolved to a relative tolerance of 
approximately $1\cdot10^{-12}$ 
\cite{mendoza_critical_phenomena_electromagnetic_2021}.
The evolutions for  $E_{\mathrm{dipole}}$ were  performed to a relative 
tolerance of only $\approx 1\cdot10^{-4}$ and for the sole purpose of 
verifying that we had initial data consistent with 
\cite{mendoza_critical_phenomena_electromagnetic_2021}.

Figure \ref{em_results_alpha_min} directly compares our simulations to those of 
\cite{mendoza_critical_phenomena_electromagnetic_2021} using both our data 
and data provided by Mendoza and Baumgarte~\cite{baumgarte_email}. This figure 
plots the minimum value of $\alpha$ on each spatial slice for family 
$E_{\mathrm{quad}}$ for marginally subcritical simulations. Comparing our 
data, we observe a significant divergence at $\tau\approx18$;  earlier than 
would be expected based on the relative precision of our searches. Similarly, 
the scaling of 
Figs.~\ref{em_results_bqp_alpha_delta}--\ref{em_results_bqp_gamma} agree with 
Figs.~2 and 7 of 
Mendoza and Baumgarte~\cite{mendoza_critical_phenomena_electromagnetic_2021} 
until $-\ln(\tau^\star-\tau)\approx0$ and $|p-p^{\star}|/p^{\star} 
\approx 1\cdot10^{-10}$ respectively. 
Past this point, the growth we observed increases relative to what was 
observed in \cite{mendoza_critical_phenomena_electromagnetic_2021}.

\begin{figure}[!ht]
\centering
\includegraphics[scale=1.0]{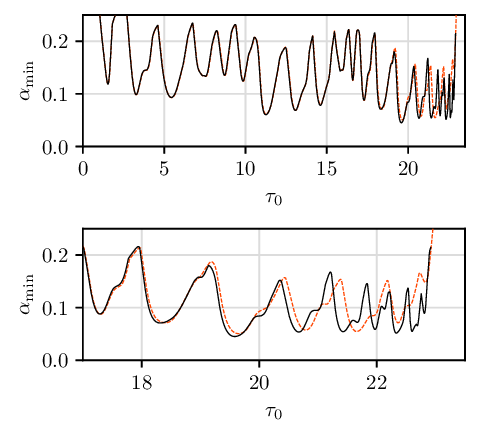}
\caption{ Minimum value of $\alpha$ on each spatial slice for family 
$E_{\mathrm{quad}}$ vs the proper time at the origin, $\tau_0$. The data 
plotted here represents the subcritical simulations closest to criticality for 
both our investigation (solid black line) and that of Mendoza and 
Baumgarte~\cite{mendoza_critical_phenomena_electromagnetic_2021, 
baumgarte_email} (dashed red line). The lower plot highlights the difference 
in behaviour at late times. Note that we have scaled $\tau_0$ for the 
data of Mendoza and Baumgarte by a factor of $\approx 1.003$ to better align 
the early minima and maxima of $\alpha_0$ with our own data. This degree of 
rescaling should be understood within the context of our simulations being 
only second order accurate and is performed to eliminate the dominant source 
of variation in our results far from the critical point. The simulations 
begin to differ markedly at $\tau \approx 18$,  earlier than would be expected 
based on the relative precision of our searches. }
\label{em_results_alpha_min}
\end{figure}

\begin{figure}[!ht]
\centering
\includegraphics[scale=1.0]{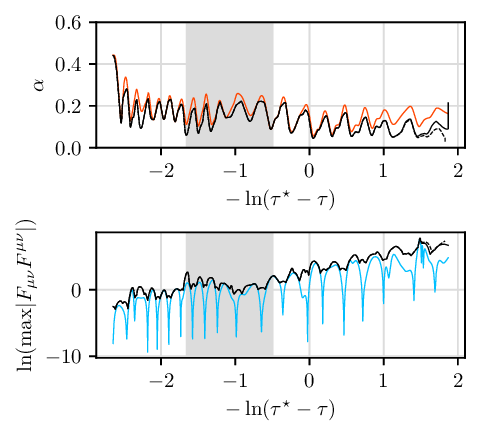}
\caption{Lapse, $\alpha$, and invariant scalar, 
$\lp|F_{\mu\nu}F^{\mu\nu}\rp|$, at the center of collapse for family  
$E_{\mathrm{quad}}$ as a function of $-\ln\lp( \tau^\star - \tau \rp)$ for 
marginally subcritical (solid black line) and supercritical (dashed black line) 
solutions with $\lp| p^\star-p\rp|/p^\star \approx 4\cdot10^{-15}$. Here, the 
black lines show the extremal values obtained on a spatial slice while the 
colored lines show the values at the center of collapse as determined by the 
coordinate location with largest value of $\lp|F_{\mu\nu}F^{\mu\nu}\rp|$ in 
the subcritical simulation closest to criticality.  }
\label{em_results_bqp_alpha_delta}
\end{figure}   

Assuming that family $E_{\mathrm{quad}}$, like family $E_{l=2}$, is best 
described by dividing the near-critical evolution into early and late time 
behaviour, a naive measurement of $\Delta$ under the assumption that our 
observer at fixed coordinate location is approximately inertial, gives 
$\Delta_1 = 0.30(5)$, $\Delta_2 = 0.33(2)$ for the first region and 
$\Delta_1 = 0.59(4)$, $\Delta_2 = 0.57(4)$ for the second region. 
Application of~(\ref{em_results_Delta_scaling}) gives $\Delta_3 = 0.61(11)$ 
for the first region and $\Delta_3 = 0.59(10)$ for the second. The large 
discrepancy between the values of $\Delta$ computed in the first region 
likely indicates that the solution does not show DSS behaviour far from 
criticality. 

Fig.~\ref{em_results_bqp_gamma} plots $\lp|F_{\mu\nu}F^{\mu\nu}\rp|$ as a 
function of $\lp|{p^\star-p}\rp|/{p^\star}$ and is used to determine 
$\gamma_{\mathrm{quad}} = 0.152(20)$. This in turn is consistent with the 
values of $\gamma$ determined for all other families. It is clear that the 
early behaviour of family $E_{\mathrm{quad}}$ is very different from that of 
family $M_{l=2}$, which indicates that the early scaling behaviour observed 
for both families may simply be the result of radiation of features of the
initial data on the path to criticality. Again we note that we list the 
complete set of $\Delta$ and $\gamma$ for family $E_{\mathrm{quad}}$ as well 
as for the families defined in Table~\ref{em_maxwell_initial_data_families} in 
Table~\ref{em_results_em_results} of the previous section. 

\begin{figure}[!ht]
\centering
\includegraphics[scale=1.0]{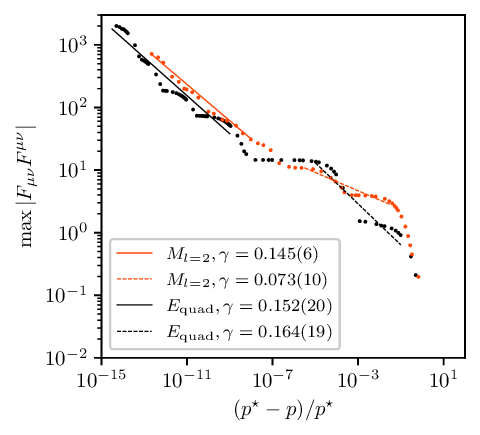}
\caption{ $\gamma$ determined via the scaling of the invariant scalar 
$\lp|F_{\mu\nu}F^{\mu\nu}\rp|$ which should scale as 
$\lp| p^\star-p \rp|^{-2\gamma}$. Plotted here are the maximum values of 
$\lp|F_{\mu\nu}F^{\mu\nu}\rp|$ obtained in each subcritical run as a 
function of $\lp|{p^\star-p}\rp|/{p^\star}$. It is apparent that although 
both quadrupole solutions exhibit scaling with similar $\gamma$ close 
to criticality the initial behaviour is highly family dependent. }
\label{em_results_bqp_gamma}
\end{figure}  

It is apparent that, close to criticality, the growth rates and echoing 
periods we observe for family $E_{\mathrm{quad}}$ differ markedly from 
those observed in~\cite{mendoza_critical_phenomena_electromagnetic_2021}.
Assuming that our results are correct, we hypothesise that the use
of spherical polar coordinates with limited resolution in 
$\theta$~\cite{mendoza_critical_phenomena_electromagnetic_2021}
may have had the inadvertent effect of leaving insufficient resolution to 
resolve dipole collapse away from the center of symmetry.
If this is the case, then it is plausible  that the growth of the dipole 
mode was suppressed in a manner similar to what is apparently observed.

\section{\label{em_sec_conclusions}Summary and Conclusions}

We have investigated the critical collapse of both the massless scalar
field and the Maxwell field in axisymmetry using the GBSSN formulation of 
general relativity.  Our study of the scalar field was largely 
motivated by the need to calibrate our numerical 
methods---including AMR---and to develop analysis procedures.  Nonetheless
we are able to reproduce previous results on massless scalar collapse to 
the estimated accuracy of our calculations.  
Moreover, in contrast to 
some other earlier work~\cite{baumgarte2018aspherical, choptuik2003_msf_axi, 
gundlach_critical_phenomena_2007}, we find no evidence of non-spherical 
unstable modes at criticality. However, as we have not examined this issue 
very closely we feel that it is well worth further study.

With regard to the Einstein-Maxwell system, we observe that for generic 
initial data a dipole mode with $\gamma_{l=1} \approx 0.149(9)$ and 
$\Delta_{l=1} \approx 0.62(8)$ seems to be dominant. If there is an
unstable quadrupolar mode, variations between the families $M_{l=2}$ 
and $E_{\mathrm{quad}}$ of Tables~\ref{em_maxwell_initial_data_families} 
and \ref{em_paper_initial_data_families} suggest that it is not universal.

We observe significant differences in the behaviour of family 
$E_{\mathrm{quad}}$ close to criticality relative to the results reported 
in~\cite{mendoza_critical_phenomena_electromagnetic_2021}, 
although our findings appear largely similar until $\lp| p^\star-p\rp|/
p^\star \approx 1\cdot10^{-10}$. 
We hypothesize that these differences may be due to the inability of spherical 
coordinates to fully resolve off-center collapse when limited angular 
resolution is employed.

The observed consistency between $\gamma$ and $\Delta$ for each of the 
families in conjunction with the observed variance in the form of $f(x)$ 
(seen in Figs.~\ref{em_results_em_gamma} and  \ref{em_results_bqp_gamma}) 
and absence of perfect DSS (seen in 
Figs.~\ref{em_results_em_m_dipole}--\ref{em_results_em_m_quad} and 
Fig.~\ref{em_results_bqp_alpha_delta}) is puzzling and requires additional 
study. Conservatively, it could be that given the slow growth rate of the 
dipolar critical solution, our simulations have simply not radiated away all 
traces of their initial data and this manifests in the apparent 
inconsistency of $f(x)$.

\begin{acknowledgments}
This research was supported by the Natural Sciences and Engineering Research 
Council of Canada (NSERC). We would additionally like to thank Maria Perez 
Mendoza and Thomas Baumgarte for generously providing their data, which was 
instrumental in our comparative analysis (see 
Sec.~\ref{em_subsec_direct_comparison} and Fig.~\ref{em_results_alpha_min}).
\end{acknowledgments}

\bibliography{em_criticality}% Produces the bibliography via BibTeX.

\end{document}